\documentclass[A4 paper,twocolumn,superscriptaddress,preprintnumbers,amsmath,amssymb]{revtex4}
\pdfoutput=1
\usepackage{graphicx}% Include figure files
\usepackage{dcolumn}% Align table columns on decimal point
\usepackage{bm}% bold math
\usepackage{multirow}
\newcommand{\micron}{{\small$\mu$}m}
\newcommand{\microns}{{\small$\mu$}m\space}

\begin{document}

\title{Controlling metamaterial resonances via dielectric and aspect ratio effects}

\author{Sher-Yi Chiam}
\affiliation{
Department of Physics, National University of Singapore, 2 Science Drive 3, Singapore 117542}
\affiliation{NUS High School of Mathematics and Science, 20 Clementi Avenue 1, Singapore 129957}
\author{Ranjan Singh}
\affiliation{
School of Electrical and Computer Engineering,
Oklahoma State University, Stillwater, Oklahoma 74078, USA}
\affiliation{
Center for Integrated Nanotechnologies, Materials Physics and Applications Division, Los Alamos National Laboratory, Los Alamos, NM 87545, USA}
\author{Weili Zhang}
\affiliation{
School of Electrical and Computer Engineering, Oklahoma State University, Stillwater, Oklahoma 74078, USA}
\author{Andrew A Bettiol}
\email{phybaa@nus.edu.sg}
\affiliation{
Department of Physics, National University of Singapore, 2 Science Drive 3, Singapore 117542}

\date{\today}% It is always \today, today,

\begin{abstract}

We study ways to enhance the sensitivity and dynamic tuning range of the fundamental inductor-capacitor ($LC$) resonance in split ring resonators (SRRs) by controlling the aspect ratio of the SRRs and their substrate thickness. We conclude that both factors can significantly affect the $LC$ resonance. We show that metafilms consisting of low height SRRs on a thin substrate are most sensitive to changes in their dielectric environment and thus show excellent potential for sensing applications.   \end{abstract}

%Use showkeys class option if keyword
\maketitle

There has been intense interest in metamaterials in the past decade, and a number of intriguing effects, such as negative refraction \cite{shelby}, electromagnetic cloaking \cite{cloak} and sub-wavelength focusing \cite{subwave} have been experimentally demonstrated. More recently, much research has been focused on demonstrating metamaterials which can be used for practical applications. For example, resonance frequency agility was demonstrated in a metafilm by tuning the conductivity of a patterned silicon substrate \cite{THzagile}. This allows for applications such as dynamically tunable notch filters. A metamaterial solid state phase modulator for terahertz radiation has also been demonstrated \cite{THzmod}. Metamaterials have also been studied for sensing applications, where the shift in the resonance frequency upon the application of a dielectric film over the metamaterial is exploited \cite{10optx_ranjan_asy,09apl_naib,oksense,debus}.

The ability to control and optimize the size of the resonance frequency shift in metamaterials can offer some advantages for practical applications. It can lead to greater sensitivity for detection purposes or to greater dynamic range for applications such as filters.  Recently, there has been some interest in fabricating metamaterials on thin membranes like Silicon Nitride \cite{10optx_papa,09apl94_peralta} and Parylene \cite{10apl96_liu}. Such flexible metafilms enable layering schemes and conformal placement onto curved surfaces, thus allowing even greater potential for applications. It is also believed that such metafilms will be more sensitive. In one case, the metamaterial showed sensitivity to the presence of a single atomic layer of graphene on its surface \cite{10optx_papa}. This sensitivity  results because thin substrates lower the effective permittivity of the supporting medium and allow simple transmission measurements due to lower loss.  There is thus strong motivation to study how a metamaterial on thin, flexible substrate differ from those on thicker substrates.

\begin{figure}[h]
\includegraphics[height=9.5cm]{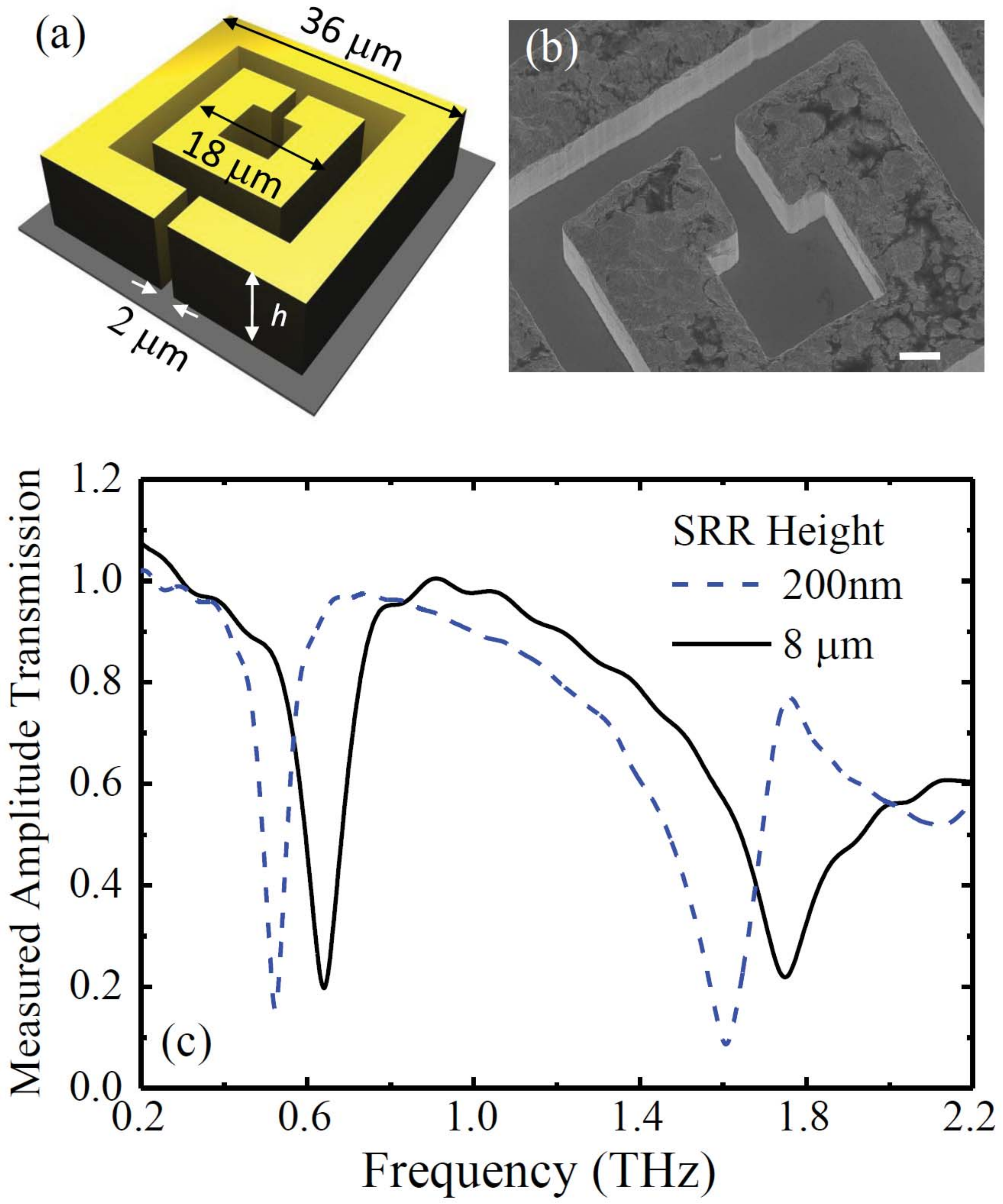}
   \caption[picnspec]
   {\label{picnspec}(a) Schematic diagram of a Split Ring Resonator (SRR) illustrating the dimensions used in this study and the SRR height parameter($h$) (b) electron micrograph of high aspect ratio SRR with $h$ = 8 \microns (scalebar = 2\micron)(c) measured THz spectra for SRRs of different heights.}
\end{figure}

In this work, we will study the effects of the substrate thickness and aspect ratio on the properties of fundamental inductor-capacitor ($LC$) resonance in the Split Ring Resonator (SRR). In particular, we want to establish if thinner substrates can indeed offer advantages like enhanced sensitivity compared to thick substrates. We will also study the effect of SRR aspect ratio on the resonance.  In a previous work, we showed the advantage of using high aspect ratio SRRs fabricated using the Proton Beam writing technique (electron micrograph in Fig. \ref{picnspec}(b)). We showed that on thick substrates, high aspect ratio SRRs offered increased shifts in the frequency of the $LC$ resonance upon the application of a dielectric layer onto the SRRs \cite{myapl}. However, it might be difficult to exploit such an advantage as there are limited facilities which can fabricate the high aspect ratio SRRs as shown. We thus also wish to establish if there is a way to increase sensitivity apart from increasing SRR height ($h$).

Figure \ref{picnspec} also shows experimental transmission spectra typical of SRRs on thick Silicon substrates (c). The SRRs were designed to have resonance frequencies in the Terahertz (THz) region and their lateral dimensions are shown in Fig. \ref{picnspec}(a)\cite{azad}. The spectra were collected using THz Time Domain Spectroscopy \cite{utopsetup,grischkowsky} for two samples of different $h$.  Measurements were done with the sample plane normal to the beam and with the THz beam polarized such that the electric field was parallel to the gap arm of the split rings.

The spectra for both samples is typical of SRRs.  We attribute the transmission dips at 0.6 THz to the so-called $LC$ resonance of the outer ring in each sample; and the dips at about 1.6 THz to an electrical dipole resonance of the outer rings. Our simulated spectra (not shown) had excellent agreement with experimental data. Simulations in this work were performed using the commercially available software, Microwave Studio\texttrademark\space from Computer Simulation Technologies. These were carried out for a single unit cell in the time domain.  We modeled gold as a lossy metal with conductivity $\sigma = 4.09 \times 10^7 Sm^{-1}$. The relative electrical permittivity ($\epsilon$) of the Si substrate is taken to be 11.6 \cite{grischkowsky} with tangent $\delta = 4 \times 10^{-3}$.

Here, we focus on the $LC$ resonance which is often exploited for functionalities such as thin film sensing \cite{oksense,myapl}. We first wish to examine the effects of the substrate thickness on this $LC$ resonance. As we can see in Fig \ref{picnspec}(c), the SRR height does also affect the $LC$ resonance. It is thus necessary to study the combined influence of both effects.

\begin{figure}[h]
\includegraphics[height=6.5cm]{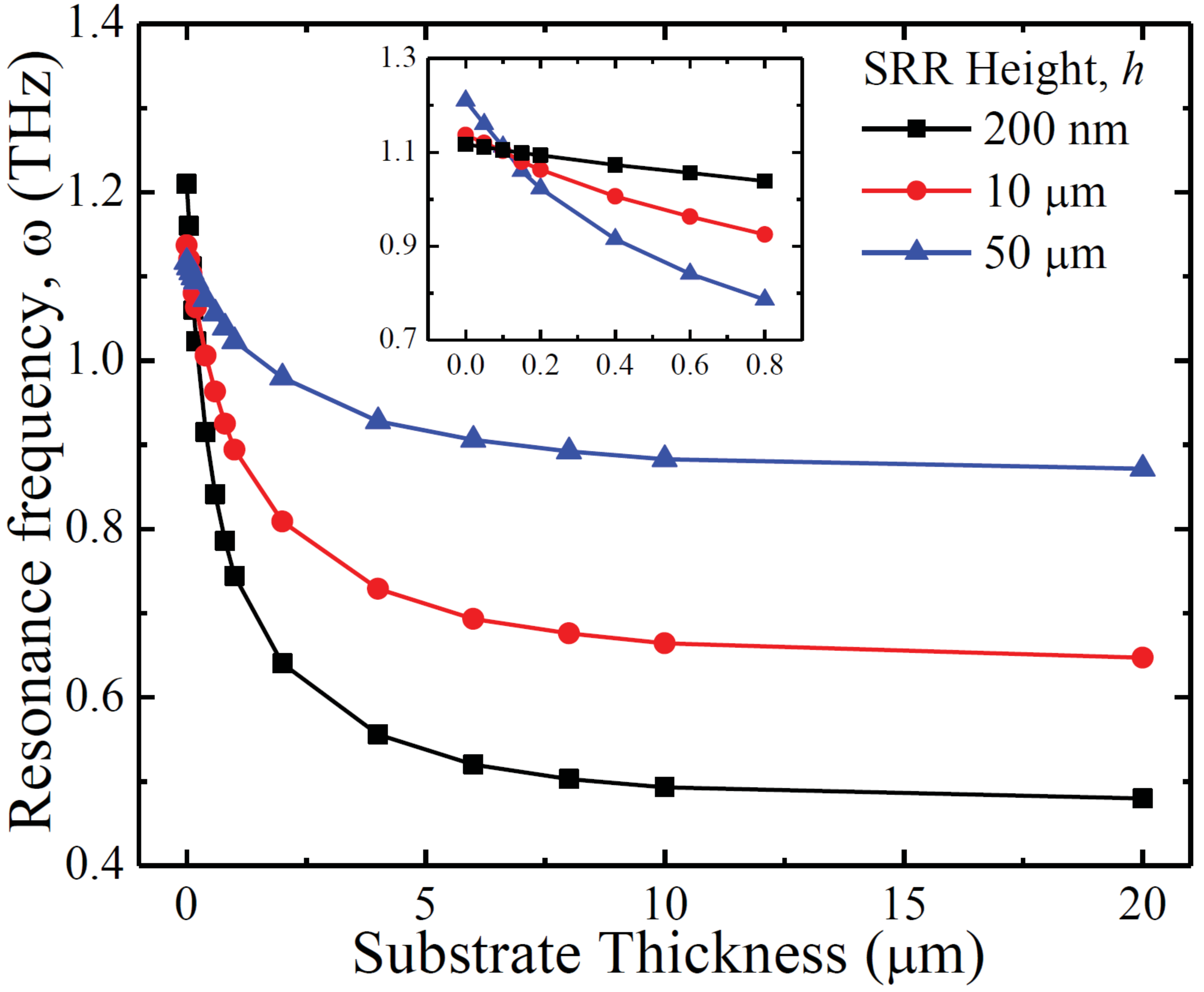}
\caption[crossover]{\label{crossover}Variation in the simulated $LC$ resonance frequency ($\omega$) as the substrate thickness is increased for SRRs of height 200 nm (black curve, squares), 10 \microns (red curve, circles) and 50 \microns (blue curve, triangles). The inset shows details of the region where the substrate is vanishingly thin, and reveals that the three curves cross. }
\end{figure}

In Fig. \ref{crossover}(a), we show the $LC$ resonance frequency ($\omega$) for SRRs of different heights as a function of substrate thickness. Here, we see that $\omega$ is most sensitive to changes in the substrate thickness when the substrate is thin for all three values of $h$. However, SRRs with the lowest height of  $h = 200 nm$ (black curve, squares) are most sensitive.  In this case, $\omega$ is just above 1.1 THz with no substrate, and drops by over 600 GHz when 5 \microns of Si substrate is added. This is a very large shift and indicates that metamaterials on substrates of sub-micron thickness would be especially sensitive even to very small changes in their dielectric environment.

It should be noted that when the substrate is sufficiently thin, $\omega$ can be shifted by adding more dielectric to the \emph{substrate} side of the metamaterial film. This effect can be exploited for sensing applications where it is not desirable to apply the analyte sample as an overlayer (i.e. from the metal film side) due to durability and compatibility issues. The thinness of the substrate allows the analyte sample film to be detected via its effect on the dielectric environment of the metafilm. In this case, the frequency shift is enhanced if low aspect height SRRs are used. When a thick substrate is present, applying the analyte firm to the substrate side would result in only a very small shift in $\omega$, even if the sample film is relatively thick (increasing substrate thickness from 10 \microns to 20 \microns leads to only an small change in $\omega$ in Fig. \ref{crossover}). Thus, for metamaterials on thick substrates, the usual practice would be to apply the analyte sample as an overlayer.

\begin{figure}[h]
\includegraphics[height=6.5cm]{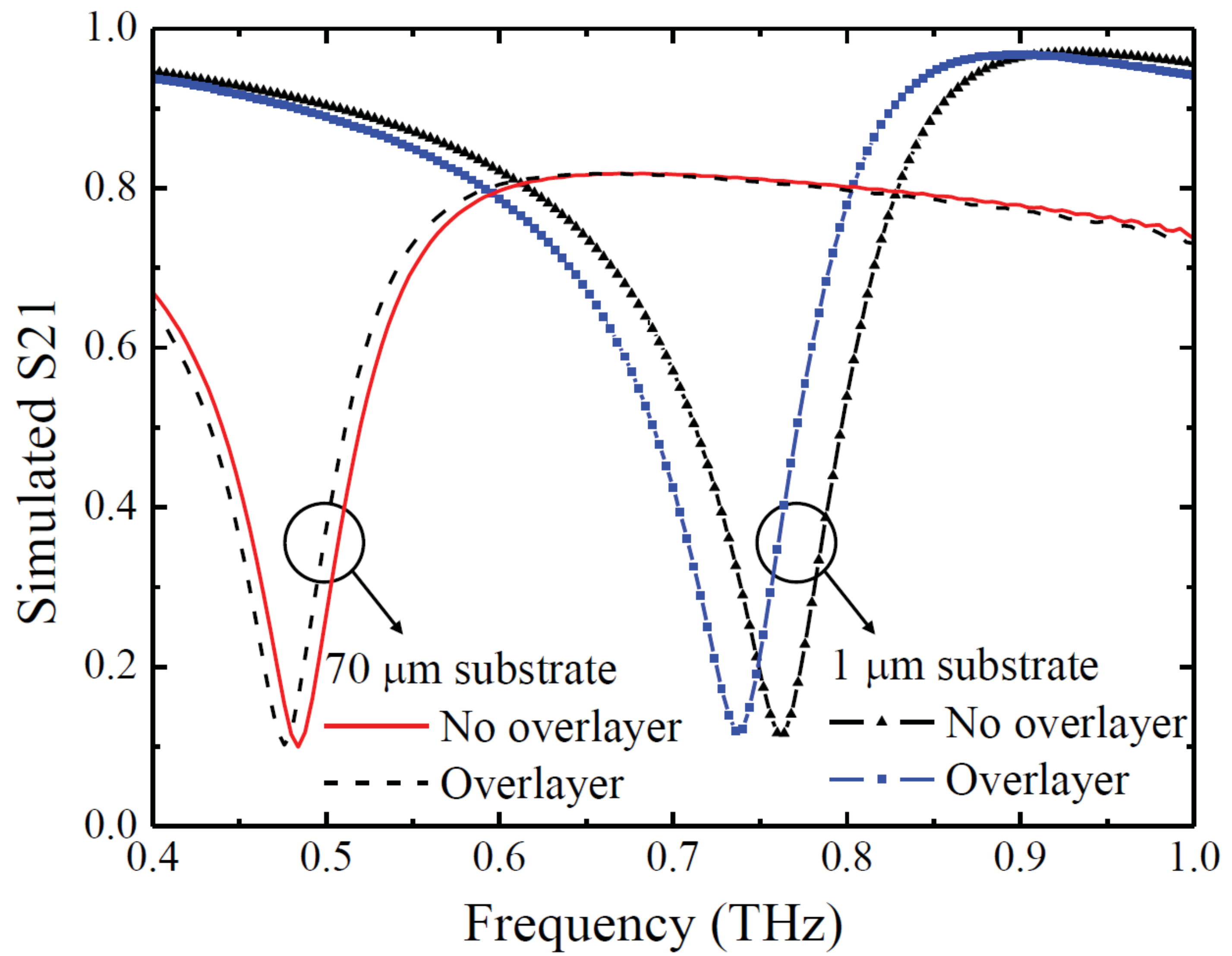}
\caption[compare]{\label{compare} Effect of substrate thickness on sensitivity : shift in $\omega$ upon addition of 1 \microns overlayer of $\epsilon$ = 2 onto SRRs on silicon substrates 1 \microns thick and 70 \microns thick.}
\end{figure}

Even when the analyte film is to be applied as an overlayer, considerable advantage can be gained with a thinner substrate. Figure \ref{compare} illustrates the effect of substrate thickness on the sensitivity of SRRs to the application of a thin dielectric overlayer. We show the effect of adding a 1 \microns thick overlayer of a dielectric with $\epsilon$ = 2.0 onto an SRR ($h$ = 200nm). The results for two substrate thickness (1 \microns and 70 \micron) are compared. We see that the frequency shift for the thinner substrate (almost 30 GHz) is much larger than the  change for the thicker substrate ($<$10 GHz). Our results thus indicate that thin substrates can significantly enhance the sensitivity of SRRs. Such an enhancement would be most pronounced if a sub-micron layer of a low dielectric constant material, such as Silicon Nitride, is used. This approach is an alternative to using high aspect ratio SRRs on thick substrates. A thinner substrate would be especially useful in cases where the sample volume is very limited, as the advantage of using high aspect ratio SRRs is most apparent only when there is sufficient sample to completely cover the SRRs \cite{myapl}.

A prominent feature of Fig \ref{crossover} is the crossing of the three curves at a substrate thickness of about 0.1 \micron. This is shown in greater detail in the inset. This shows that the substrate thickness can influence the $\omega$-$h$ relationship. If the substrate is thinner than 0.1 \microns (or totally absent), $\omega$ decreases with $h$.  If it is thicker, $\omega$ increases with $h$ instead.

The latter trend is evident from the experimental data in Fig. \ref{picnspec}, where the SRRs are on thick Si substrates. Fig.\ref{picnspec}(c) shows that $\omega$ for $h$ = 8 \microns (black solid curve) is higher than for $h$ = 200 nm (blue dashed curve). This is consistent with results from other experimental work \cite{guo,okthickness}. Situations in which $\omega$ decreases with $h$, are however, not as well studied or understood. We wish to study this effect in greater detail here.

\begin{figure}[h!]
\includegraphics[height=3.5cm]{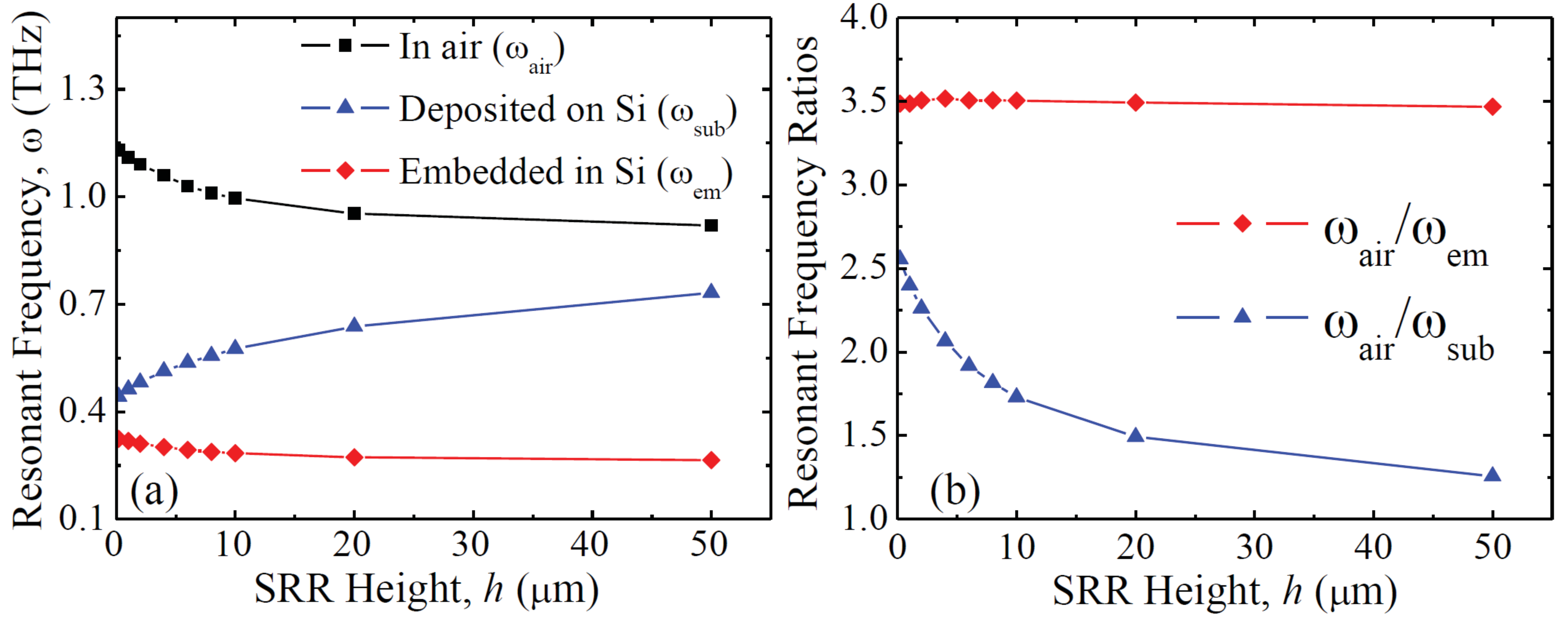}
\caption[ratios]{\label{ratios}Plot of (a) $LC$ resonance frequencies in three different dielectric environments (b) frequency ratios against SRR height.}
\end{figure}

In Fig. \ref{ratios}(a), we show simulated data which tracks how $\omega$ varies as a function of $h$ for SRRs in three different dielectric environments: freely suspended in air ($\omega_{air}$ - black curve with squares), deposited on a thick Si substrate ($\omega_{sub}$ - blue curve with triangles) and completely embedded in Si ($\omega_{em}$ - red curve with diamonds). We also show a graph of $\frac{\omega_{air}}{\omega_{em}}$ and $\frac{\omega_{air}}{\omega_{sub}}$ against $h$ in Fig. \ref{ratios}(b).

Here, we see that $\omega_{air}$ and $\omega_{em}$ decrease with $h$ in the same way. We see from Fig. \ref{ratios}(b) that for all values of $h$, $\frac{\omega_{air}}{\omega_{em}}$ (red curve with diamonds) is almost constant at about 3.5. This value is close to the squareroot of the relative electric permittivity used for Si ($\epsilon_{Si}=11.6$) in the simulations ($\sqrt{11.6} = 3.41$). This indicates that if an SRR is located within an uniform dielectric medium, $\omega$ decreases with $h$. The dielectric shifts the $LC$ resonance frequency by a constant factor equal to $\sqrt{\epsilon}$ regardless of $h$. This is expected because the capacitance of the SRR is proportional to $\epsilon$.

From Fig. \ref{ratios}(b), we see that $\frac{\omega_{air}}{\omega_{sub}}$ (blue curve with triangles) is always greater than unity. This indicates that the substrate lowers the $LC$ resonance frequency, even if it is not actually present in the SRR gap.  The substrate must therefore increase the capacitance of the SRR. This happens because of the fringing electric field which extend beyond the SRR gap into the substrate. The importance of these fringing field has been experimentally shown \cite{10optx_shelton}. In addition, the ratio $\frac{\omega_{air}}{\omega_{sub}}$ is not constant but decreases with $h$. This is to be expected. The gaps of low-height SRRs resemble strip capacitors where the fringing field in Si contribute significantly to the capacitance. The Si substrate influences the capacitance considerably and lowers $\omega_{sub}$ by a large factor ($\frac{\omega_{air}}{\omega_{sub}}$ = 2.56 for $h$ = 200nm). As $h$ increases, the gap begins to resemble a parallel plate capacitor. The air-filled gap dominates the capacitance of the SRR gap and the fringe field in the substrate became less important. Furthermore, the upper regions of the SRR are at a distance from the Si substrate and thus are not sensitive to its presence.  Thus $\omega_{sub}$ should start to approach $\omega_{air}$ for very tall SRRs ($\frac{\omega_{air}}{\omega_{sub}}$ =1.26 for $h$ = 50\micron).

The results of this analysis thus gives us a more complete understanding of the effects of the substrate and SRR aspect ratio on the resonance of SRRs. We can deduce that when SRRs are in an uniform dielectric environment, $\omega$ decreases as SRR height increases. When the SRR is located on a substrate, the dielectric effect of the substrate is a function of SRR height.

In conclusion, we have studied the influence of aspect ratio and substrate dielectric constant in detail for SRRs in the THz regime. In this process, we gained a better understanding of the dual effects of substrate and SRR aspect ratio and showed that these factors can be used as a means to control metamaterial resonances and thus to enhance performance for sensing applications. Our results indicate that metafilms on thin substrates can offer increased sensitivity through a larger frequency shift upon the application of an ultrathin dielectric film.

The work was funded partially by the National University of Singapore grant NUS R144 000 204 646 and the U.S. National Science Foundation Grant No. ECCS-0725764.

%\bibliography{meta1}
%\bibliographystyle{apsrev}

\end{document}